\documentclass[eqsecnum,aps,superscriptaddress,11pt ]{revtex4}
\usepackage{graphicx}
\usepackage{dcolumn}
\usepackage{xspace}

\begin{document}
\title{{\it Ab-initio} relativistic many-body calculation of hyperfine splittings of $^{113}$Cd$^{+}$ }
\vspace*{0.5cm}

\author{Gopal Dixit$^1$, H. S. Nataraj$^2$, B. K. Sahoo$^3$, R. K. Chaudhuri$^2$ and Sonjoy Majumder$^1$ \\
\vspace{0.3cm}
{\it $^1$Department of Physics, Indian Institute of Technology-Madras, Chennai-600 036, India} \\
{\it $^3$ Indian Institute of Astrophysics, Bangalore-34, India}\\
{\it $^2$ Max Planck Institute for the Physics of Complex Systems, D-01187 Dresden, Germany}\\
}
\date{\today}

\begin{abstract}
\noindent
This work presents  accurate  {\it ab initio} determination of the hyperfine splitting for the ground state and a few low-lying excited 
states of  $^{113}$Cd$^{+}$; important candidate for the frequency standard in the microwave region, using  coupled-cluster theory (CC) 
in the relativistic framework. The hyperfine energy splittings, calculated first time in literature, are in good agreement with the recent experimental results. We 
have also carried out the lifetimes of the $5p\,^{2}P_{1/2}$ and $5p\,^{2}P_{3/2}$ states, which are in good agreement with the available 
experimental results. The role of different electron correlation effects in the determination of these quantities are discussed and their 
contributions are presente in the CC terms.
\end{abstract}
\maketitle

\section{Introduction}
The current frequency standard is based on the ground state hyperfine transition in $^{133}Cs$ which is in the microwave regime and has an uncertainty of one part in $10^{15}$ \cite{atom}. Trapped and laser cooled ions are excellent candidates for many high precision measurements \cite{ghosh,itano}. Due to the decoupling of the internal states caused by the perturbations arising from the collisions and Doppler shifts, trapped and laser-cooled ions have been regarded as nearly isolated quantum systems \cite{tanaka}. Therefore, precise measurement of transition frequencies of several species of trapped ions have been performed for the purpose of developing the better frequency standard in microwave and optical frequency regions \cite{tanaka,u.tanaka}. Singly ionized cadmium (Cd$^{+}$) has a  potential for the applications in quantum information processing where the  microwave transition between the hyperfine states of the ground state is used for both a re-pumping process and manipulation of quantum states of trapped ions \cite{tanaka}. Recently, $^{113}$Cd$^{+}$ has been proposed for the design of a space qualified atomic clock \cite{jelenkovic}. In this respect, a series of measurements for the ground state hyperfine splitting of $^{113}$Cd$^{+}$ have been performed \cite{u.tanaka,tanaka,jelenkovic}.  In recent experiments, the control of quantum states of $^{113}$Cd$^{+}$ with high degree has become possible \cite{blinov} because it has a simple energy-level structure and an accessible wavelengths for excitations. This can also be regarded as a step towards the development of a new frequency standard in the microwave regime \cite{tanaka}. In this work, we have used the coupled-cluster theory with single, double and partial triple excitations (CCSD(T)) in the relativistic frame work  to calculate the  hyperfine splitting of ground state and few low-lying excited states of $^{113}$Cd$^{+}$. This is the first {\it ab-initio} relativistic many-body study of the hyperfine splitting of $^{113}$Cd$^{+}$ in literature, known to our knowledge.  \\


The hyperfine interaction in an atom is generated due to the interaction of different electromagnetic multipole moments of the nucleus and of an atom. The corresponding Hamiltonian is given by \cite{Cheng}
\begin{equation}
H_{hfs} = \sum_k \bf{M}^{(k)}\cdot \bf{T}^{(k)},
\end{equation}
where $\bf{M}^{(k)}$ and $\bf{T}^{(k)}$ are the spherical tensor operators of
rank $k$ of the nucleus and the electronic system, respectively.
Since, nuclear spin of $^{113}$Cd$^+$ is $1\over 2$, hyperfine splitting due to electric quadrupole moment
will be zero. Magnetic hyerfine interaction, which is the only important
hyperfine interaction in this system, for a relativistic electron with nuclear magnetic
moment $\mu_I$  is given by \cite{tpdas}
\begin{equation}
H_{hfs} = \sum_i e c{\bf {\alpha}_i}\cdot \frac{\bf {\mu}_I\times {\bf r}_i}{r_i^3}
\end{equation}
${\bf \alpha_i}$ are Dirac matrix for the $i$th electron.

In the first-order perturbation theory, the hyperfine energies $E_{hfs}(J)$ of the
fine-structure state $|JM_J\rangle$ are the expectation values of the hyperfine interaction Hamiltonian.
Details of the expression are given in Cheng and Childs \cite{Cheng}. The energies corresponding to
magnetic dipole hyperfine transitions are defined as
\begin{equation}
E_{M1}=AK/2
\end{equation}
Here $K=2\langle I \cdot J \rangle = F(F+1)-I(I+1)-J(J+1)$ with $I$ and $J$ are the total angular momentum of the
nucleus and the electronic state, respectively. $F$ is the total angular momentum of an atom (nucleus + electron).
The hyperfine constants $A$  corresponding to the magnetic dipole hyperfine interactions are given by
\begin{equation}
A={1\over {IJ}}\langle J|H_{hfs}|J\rangle = \mu_Ng_I\frac{\langle J ||T^{(1)}||J\rangle}{\sqrt{J(J+1)(2J+1)}},
\end{equation}
where $\mu_N$ is the Bohr magneton, $g_I=\mu_I/I$ where $\mu_I$ is the nuclear dipole moment. The $T^{(1)}$ operator is defined as
\begin{equation}
T^{(1)} = \sum_i -ie\sqrt{8\pi/3}r_i^{-2} {\bf \alpha_i}\cdot {\bf Y}_1^{(0)}\bigl(\hat{r_i}\bigr)
\end{equation}
with ${\bf Y}_1^{(0)}$ as vector spherical harmonic.

The Fock-space Multi-reference  Coupled Cluster (FSMRCC) theory for one electron attachment process used here has been described elsewhere \cite{lindgren,mukherjee,Haque,Pal}.
We provide a brief review of this method. The theory for a single valence system is based on the
concept of common vacuum for both the closed shell N and open shell N$\pm$1 electron systems, which
allows us to formulate a direct method for energy differences. Also,
the holes and particles are defined with respect to the common vacuum for
both the electron systems. Model space of a (n,m) Fock-space
contains determinants with $n$ holes and $m$ particles distributed within
a set of what are termed as {\em active} orbitals. For example, in this
present article, we are dealing with (0,1) Fock-space which is a complete
model space (CMS) by construction and is given by

\begin{equation}\label{eq4}
|\Psi^{(0,1)}_\mu\rangle=\sum_i {\mbox C}_{i\mu} |\Phi_i^{(0,1)}\rangle
\end{equation}
\noindent
where ${\mbox C}_{i\mu}$'s are the coefficients of $\Psi^{(0,1)}_\mu$
and $\Phi^{(0,1)}_i$'s are the model space configurations.
The dynamical electron correlation effects are introduced through the
{\em valence-universal} wave-operator $\Omega$
\cite{lindgren,mukherjee}
\begin{equation}\label{eq5}
\Omega={\{ \exp({\tilde{S}}) }\}
\end{equation}
\noindent
where
\begin{equation}\label{eq6}
{\tilde{S}}=\sum_{k=0}^m\sum_{l=0}^n{S}^{(k,l)}
                 ={S}^{(0,0)}+{S}^{(0,1)}+ {S}^{(1,0)}+\cdots
\end{equation}
At this juncture, it is convenient to single out the core-cluster
amplitudes $S^{(0,0)}$ and call them $T$. The rest of the
cluster amplitudes will henceforth be called $S$. Since $\Omega$
is in normal order, we can rewrite Eq.(\ref{eq5}) as
\begin{equation}\label{eq7}
\Omega ={exp(T)}{\{\mbox{exp}({S}) }\}
\end{equation}

In this work, single ($T_{1}, S_{1}$) and double excitations ($T_{2}, S_{2}$) are considered for $T$ and $S$  clusters operator.
Wavefunction of the system with single valence orbital {\it v}
\begin{equation}
|\Psi_v\rangle = \Omega_{v}\Phi_{DF}\rangle = e^{T_{1}+T_{2}}\{1+S_{1v}+S_{2v}\}|\Phi_{DF}\rangle.
\end{equation}
Triple excitations are included in the open shell CC amplitude which
correspond to the correlation to the valence orbitals, by an approximation
that is similar in spirit to CCSD(T) \cite{ccsd(t)}. The approximate
valence triple excitation amplitude is given by

\begin{equation}
{S^{(0,1)}}_{abk}^{pqr}=\frac{{\{{\overbrace{V{T}_2}}\}_{abk}^{pqr}}+{\{{\overbrace{V{S^{(0,1)}}_2}}}\}_{abk}^{pqr}}{\varepsilon_{a}+\varepsilon_{b}+\varepsilon_{k}-\varepsilon_{p}-\varepsilon_{q}-\varepsilon_{r}},\label{eq21}
\end{equation}
where ${S^{(0,1)}}_{abk}^{pqr}$ are the amplitudes corresponding to the simultaneous
excitation of orbitals $a,b,k$ to $p,q,r$, respectively;
$\overbrace{V{T}_2}$ and $\overbrace{V{\mbox S^{(0,1)}}_2}$ are the connected
composites involving $V$ and $T$, and $V$ and $S^{(0,1)}$, respectively, where
$V$ is the two electron Coulomb integral and $\varepsilon$'s are the
orbital energies.

The expectation value of any operator $O$ can be expressed, in the CC method, as
\begin{eqnarray}
O& = & \frac{\langle \Psi_v|O|\Psi_v\rangle}{\langle \Psi_v|\Psi_v\rangle} \nonumber \\
&=& \frac{\langle \Phi_v|\{1+{S_v}^{\dag}\}{e^T}^{\dag}Oe^T\{1+S_v\}|\Phi_v\rangle}
{\langle \Phi_v|\{1+{S_v}^{\dag}\}{e^T}^{\dag}e^T\{1+S_v\}|\Phi_v\rangle}
\end{eqnarray}

The contribution from the normalization factor is given by,
\begin{eqnarray}
Norm = \langle \Psi_v | O | \Psi_v \rangle \{ \frac {1}{N_v} - 1 \}
\end{eqnarray}
with $N_v = \langle \Phi_v | e^{T^\dagger} e^T + \{ S_v^{\dagger} e^{T^{\dagger}} e^T S_v \} | \Phi_v \rangle$ for the valence electron $v$.

We calculate the  DF wavefunctions $|\Phi{_{DF}}\rangle$ using the Gaussian-type orbitals (GTO) as given in \cite{rajat} using basis function of the form \\
\begin{equation}
F^{L/S}_{i,k}(r) = C_N^{L/S}r^{k}e^{-\alpha_{i}r^2}
\end{equation}
with $k=0, 1, 2, 3,....$ for s, p, d, f ..... type orbital symmetry respectively \cite{rkc}. The radial functions `$F^{L}$' and `$F^{S}$'
represent the basis functions correspond to large and small components of the Dirac orbitals.
$ C_N^{L/S}$ are the normalization constant which depend on the exponents. The universal even tempering condition has been applied to the exponents
; i.e., for each symmetry exponents are assigned as
\begin{equation}
\alpha_{i}=\alpha_0\beta^{i-1} \hspace{1in} i=1,2,.....N
\end{equation}
where N is the number of basis functions for the specific symmetry.
In this calculation, we have used $\alpha_0=0.00525$ and $\beta=2.73$. The number of basis functions used in the present calculation is
32, 32, 30, 25, 20, 20 for $l=$ 0, 1, 2, 3, 4 symmetries, respectively.

Number of DF orbitals for different symmetries used in the CC calculations
are based on convergent criteria of core correlation energy for which it satisfies numerical completeness.
There are only 10, 9, 8, 7 and 5 active orbitals including all core electrons are considered in the CCSD(T) calculations
for $l=$ 0, 1, 2, 3, 4 symmetries, respectively. We first calculate $T$ amplitudes using the CC equations of closed
shell systems and then solve the $S$ amplitudes from the open shell equation for this single valence states of Cd$^{+}$.

\begin{table}[h]
\caption{Radiative lifetime(ns) for different low-lying states of $^{113}$Cd$^{+}$.}
\begin{ruledtabular}
\begin{tabular}{lrrr}
State &  Experiment & Other theories & This work   \\
\hline
${^2}P_{1/2}$  &  3.2(2),$^{a}$ 3.05(13),$^{b}$ 3.11(4),$^{c}$
               & 2.92,$^{a}$ 2.99,$^{a}$ 2.74,$^{b}$  & 3.093 \\
               & 3.5(2),$^{c}$ 4.8(1.0),$^{d}$ 3.14(0.011)$^{e}$ & 3.11$^{f}$ &  \\
${^2}P_{3/2}$  &  2.5(3),$^{a}$ 2.70(25),$^{b}$ 2.77(7),$^{c}$   & 2.50,$^{a}$ 2.3,$^{b}$   & 2.602\\
              &   3.5(2),$^{c}$  3.4(7),$^{d}$ 2.647(0.01)$^{e}$ 3.0(2),$^{g}$ & 2.77$^{f}$ & \\
\end{tabular}
\end{ruledtabular}
\label{tab:results1}
\end{table}
$^{a}$Time-resolved laser-induced fluorescence \cite{Xu}\\
$^{b}$Hanle-theory \cite{hamel} \\
$^{c}$Beam-laser, beam foil (ANDC) \cite{pinnington} \\
$^{d}$Phase-shift method \cite{baumann} \\
$^{e}$Ultrafast laser pulses \cite{moehring} \\
$^{f}$Many-body third order perturbation theory \cite{chou} \\
$^{g}$Hanle \cite{andersen} \\

We report our lifetime results along with the other calculated and measured results in table I.
As seen, there are large disagreements among earlier estimations. The most reliable experimental results are presented by Moehring et al.
\cite{moehring} to date with a total uncertainty of 0.4$\%$.  We can see from table I, our calculated lifetime for 5$P$ fine structure states of $^{113}Cd^{+}$ are in excellent agreement with the recent measured result \cite{moehring} and highly correlated MBPT calculations \cite{chou}.   \\

Table II presents the computed values of the hyperfine energy splittings for ground state and  few low-lying excited states of $^{113}Cd^{+}$.  We have used the expression (1.3) to compute the highly sensitive property to the electronic wavefunction near to the nuclear region. To calculate  the hyperfine splitting constants corresponding to the magnetic dipole $A$ with  $\mu_I$ = -0.8278  \cite{raghavan}.  It is clear from the  table II that our calculated hyperfine energy splittings are in excellent agreement with the measured hyperfine energy splittings wherever available. We have also estimated the hyperfine splitting of few other excited states for which there are no experimental results available to the best of our knowledge.   \\

All the core orbitals are considered as active in our calculations. In Table III, the individual contribution from the one body and effective two body terms of the magnetic dipole hyperfine structure constant A for $^{113}Cd^{+}$ are listed.  The first term ($O$) is the Dirac-Fock (DF) contribution. From the differences of DF and total CC results, it is evident that, the electron correlation effects to the calculated $A$ value vary from (15-45)$\%$ among different low-lying states.   \\

We know, the Brueckner pair-correlation effects are in the form of $\overline{O}\, S_{1v}$ and its conjugate terms whereas core-polarization effects are in $\overline{O}\, S_{2v}$ and its conjugate terms, in its lowest order. Both the correlations are important in the precise determination of the final results. From table III, it is clear that the largest contribution of electron correlation to the hyperfine splitting constants A for different low-lying states comes from the pair-correlation effects. However, the core-polarization contributions are not that least significant. 
In figure 1, we have plotted the important correlation effects to the hyperfine constants of different states with respect to their DF contributions. The core-correlation effects seem to be relatively small compared to the other two effects presented in the figure. The percentage contribution of the former one is almost the same for all the considered states. We see the ratio of pair-correlation and core-polarisation effects is almost one for 6$S_{1/2}$ state, which is different for other states. The pair-correlation effects for the fine structure state of 5$P$ are extremely strong (almost 25\%) with respect to DF value, whereas for other states it is around 10\%. \\
Among the other correlation effects, the prominent contributions are observed from $S_{2v}^{\dagger} \,\overline{O}\, S_{2v} + cc $, which is almost 2\% for the $S$ and $P_{1/2}$ states, and more than 5\% for $P_{3/2}$ states. Contributions from the effective two-body terms are also significant and comparable with some of the one body effects like $S_{1v}^{\dagger} \,\overline{O}\, S_{2v} + cc $ for all the states. \\ 
\begin{table}
\caption{Hyperfine energy splitting of different low-lying states of
 $^{113}$Cd$^{+}$.}
\begin{ruledtabular}
\begin{tabular}{lrr}
State & Experimental & This work    \\  \hline
5$S_{1/2}$ & 15.2(2 Hz) GHz \cite{tanaka,jelenkovic} & 15.28 GHz    \\
5$P_{1/2}$ & 2.45 GHz \cite{tanaka} & 2.43 GHz    \\
5$P_{3/2}$ & 800 MHz \cite{tanaka} & 812.04 MHz   \\
6$S_{1/2}$ &     & 3.23 GHz \\
6$P_{1/2}$ &     & 667.81 MHz \\
6$P_{3/2}$ &     & 236.24 MHz \\
\end{tabular}
\end{ruledtabular}
\label{tab:results1}
\end{table}

\begin{table}[h]
\caption{Contributions of different coupled-cluster terms to the $^{113}$Cd$^+$
 magnetic dipole ($A$) hyperfine constant.  $CC$ stands for the complex conjugate part of the corresponding terms.}
\begin{tabular}{lcccccc}
\hline
\hline
Terms & 5s$_{1/2}$ & 5p$_{1/2}$ & 5p$_{3/2}$ & 6s$_{1/2}$ & 6p$_{1/2}$ & 6p$_{3/2}$ \\
      &  state & state & state & state & state & state \\
\hline
 & & & & & & \\
  O & -11986.36 & -1837.36 & -284.56 & -2753.76 & -553.68 & -87.78 \\
$\overline{O}$ & -11896.12 & -1823.30 & -284.76 & -2734.74 & -550.04 & -87.82 \\
$\overline{O}\, S_{1v} + cc $ & -2266.42 & -476.44 & -74.19 & -248.89 & -75.64 & -12.52 \\
$\overline{O}\, S_{2v} + cc $ & -1067.66 & -107.83 & -27.36 & -223.06 & -38.22 & -9.47 \\
$S_{1v}^{\dagger} \,\overline{O}\, S_{1v}$ & -107.95  & -31.23 & -4.85 & -5.66 & -2.65 & -0.45  \\
$S_{1v}^{\dagger} \,\overline{O}\, S_{2v} + cc $ & -83.28 & -16.14  & -3.95 & -3.99 & -3.16 & -0.81 \\
$S_{2v}^{\dagger} \,\overline{O}\, S_{2v} + cc $ & -338.50 & -45.99 & -21.23 & -81.69 & -11.16 & -9.13 \\
\hline\\
\multicolumn{5}{c}{\textbf{Important effective two-body terms of $\overline{O}$ }} \\
\hline
 & & &  \\
$S_{2v}^{\dagger}\, O\, T_1 + cc $ & -65.28 & -9.57 & -1.43 & -15.64 & -2.82 & -0.43 \\
$S_{2v}^{\dagger}\, O\, T_2 + cc $ & 156.55 & 23.20 & 3.25  & 33.77 & 5.79 & 0.80 \\
 Norm. & 374.86 & 51.78 & 8.36  & 42.14 & 10.00 & 1.73 \\
Total & -15285.99 & -2434.01 & -406.02 & -3237.10 & -667.81 & -118.12 \\
\hline
\hline
\end{tabular}
\label{tab:front3}
\end{table}

\begin{figure}
\begin{center}
\includegraphics[width=2.5in]{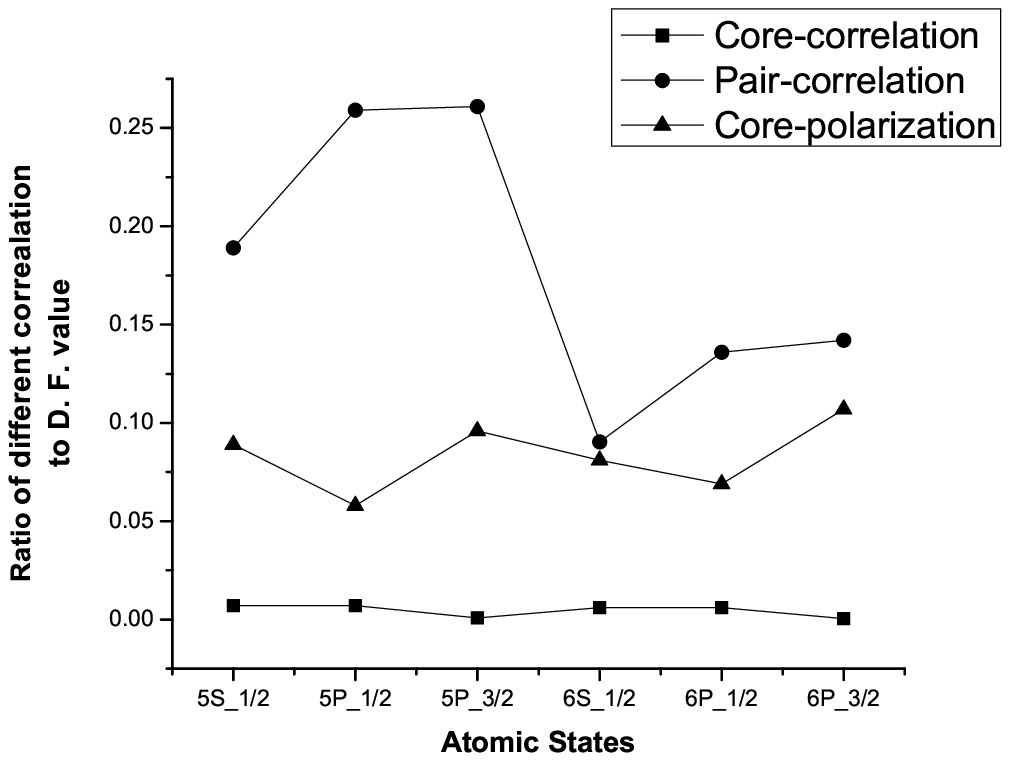}
\caption{\label{label}The ratio of core-correlation, pair-correlation and core-polarization effects w.r.t. the DF values}
\end{center}
\end{figure}

\section{Conclusion}
In this work, we have determined the hyperfine energy splittings of the ground and a few low-lying excited states
of $^{133}$Cd$^+$ using the coupled-cluster theory in relativistic framework.
Lifetimes of 5$P$ fine structure states of system are estimated also.
Our results are in excellent agreement with the available  measurements. This suggests the robustness of the CC 
method and our numerical approach in obtaining the accurate wavefunctions of the system considered. 

\section{Acknowledgment}
We are thankful to  Prof. B. P. Das, Indian institute of Astrophysics, Bangalore and Prof. Debashis Mukherjee, Indian Association of Cultivation for Science, Kolkata for helpful discussions.


\begin{thebibliography}{14}
\bibitem{atom}
http://tf.nist.gov/cesium/atomichistory.htm.
\bibitem{ghosh}
P. K. Ghosh,Ion Traps: Oxford Science Publications (Oxford: Clarendon)  (1995).
\bibitem{itano}
D. J. Wineland and W. M. Itano, Phys. Today {\bf 40}, 34  (1987).
\bibitem{tanaka}
U. Tanaka et al., Phys. Rev. A. {\bf 53}, 3982 (1996).
\bibitem{u.tanaka}
U. Tanaka et al., Appl. Phys. B. {\bf 78}, 43-47 (2004).
\bibitem{jelenkovic}
B. M. Jelenkovic et al., Phys. Rev. A.  {\bf 74}, 022505 (2006).
\bibitem{blinov}
B.B. Blinov, D.L. Moehring, L.M. Duan and C. Monroe, Nature (London) {\bf 428}, 153 (2004).
\bibitem{Cheng}
K. T. Cheng and W. J. Childs, Phys. Rev. A.  {\bf 31}, 2775 (1985).
\bibitem{tpdas}
T. P. Das, Hyperfine Interaction. {\bf34}, 189 (1987).
\bibitem{lindgren}
I. Lindgren and J. Morrison, Atomic Many-body Theory {\bf 3}, Ed. G. E.
Lambropoulos and H. Walther (Berlin: Springer) (1985).
\bibitem{mukherjee}
I. Lindgren, D. Mukherjee, \emph{Phys. Rep.} \textbf{151}, 93 (1987).
\bibitem{Haque}
A. Haque, D. Mukherjee, J. Chem. Phys. {\bf 80}, 5058 (1984).
\bibitem{Pal} 
S. Pal, M. Rittby, R. J. Bartlett, D. Sinha, D. Mukherjee,
Chem. Phys. Lett. \textbf{137}, 273 (1987); J. Chem. Phys. {\bf 88}, 4357
(1988).
\bibitem{ccsd(t)}
K. Raghavachari, G. W. Trucks, J. A. Pople, M. Head-Gordon, \emph{Chem. Phys.
  Lett.}, \textbf{157}, 479 (1989); M. Urban, J. Noga, S. J. Cole and R. J.
  Bartlett, {\bf 83}, 4041 (1985).

\bibitem{rajat}
R. K. Chaudhuri, P. K. Panda and B. P. Das, Phys. Rev. A {\bf 59}, 1187 (1999).
\bibitem{rkc}
R. K. Chaudhari, P. K. Panda, B. P. Das, U. S. Mahapatra and D. Mukherjee, J. Phys. B. {\bf 33}, 5129 (2000).
\bibitem{Xu}
H. L. Xu , A. Persson, S. Svanberg, K. Blagoev, G. Malcheva, V. Pentchev, E. Biemont, J. Campos, M. Ortiz  and R. Mayo, Phys. Rev. A.  {\bf 70}, 042508 (2004).
\bibitem{hamel}
J. Hamel  and J. P. Barrat, Opt. Commun. {\bf 10}, 331 (1974).
\bibitem{pinnington}
E. H. Pinnington, J. J. Van Hunen, R. N. Gosselin, B. Guo  and R. W. Berends, Phys. Scr. {\bf 49}, 331 (1994).
\bibitem{baumann}
S. R. Baumann and W. H. Smith, J. Opt. Soc. Am. {\bf 60}, 345 (1970).
\bibitem{moehring}
D. L. Moehring, B. B. Blinov, D. W. Gidley, R. N. Kohn, M. J. Madsen, T. D. Sanderson, R. S. Vallery  and C. Monroe, Phys. Rev. A. {\bf 73}, 023413 (2006).
\bibitem{chou}
U. I. Safronova, I. M. Savukov, M. S. Safronova and W. R. Johnson, Phys. Rev. A. {\bf 68}, 062505 (2003).
\bibitem{andersen}
T. Andersen, O. Poulsen  and P. S. Ramanujam, J. Quant. Spectrosc. Radiat. Transf. {\bf 16}, 521 (1976).
\bibitem{raghavan}
P. Raghavan, At. Data Nucl. Data tables  {\bf 42}, 189 (1989)
\end{thebibliography}
\end{document}